\documentclass[aps,prb,twocolumn]{revtex4-1}
\usepackage{graphicx}
\usepackage{amssymb}%
\usepackage{epsfig}
\usepackage{amsmath}
\usepackage{ulem}
\usepackage{color}
\usepackage{subfigure}

\begin{document}
\title{Multiple time scales from hard local constraints: Glassiness without disorder}
\author{O.~C\'epas}
\affiliation{Institut N\'eel, CNRS et Universit\'e Joseph Fourier, BP 166, F-38042 Grenoble 9, France }

\begin{abstract}
While multiple time scales generally arise in the dynamics of
disordered systems, we find multiple time scales in the absence of
disorder in a simple model with hard local constraints. The dynamics
of the model, which consists of local collective rearrangements of
various scales, is not determined by the smallest scale but by a
length $l^*$ that grows at low energies. In real space we find a
hierarchy of fast and slow regions: Each slow region is geometrically
insulated from all faster degrees of freedom, which are localized in
fast pockets below percolation thresholds. A tentative analogy with
structural glasses is given, which attributes the slowing down of the
dynamics to the growing size of mobile elementary excitations, rather
than to the size of some domains.

\end{abstract}

\pacs{PACS numbers:}
\maketitle

\section{Introduction}

Local collective rearrangements of molecules may play an important
role in the dynamics of supercooled liquids and be at the origin of
the slowing down of the relaxation time, leading to the glass
transition.\cite{AG,Stillingerreview} The microscopic nature of them
remains, however, elusive. The Adam-Gibbs interpretation\cite{AG} of
the slowing down involves the existence of ``cooperatively rearranging
regions'' which progressively grow as the glass transition is
approached, and argues for a connection between dynamics and
thermodynamics. The limitations of the interpretation are clear. The
analysis does not specify what the ``cooperatively rearranging
regions'' are. It assumes that they are independent, so that the
relaxation of the system occurs on the fastest time scale available,
given by the smallest region. It further assumes a growing length
scale and an activated behavior for the collective rearrangements,
resulting in growing activation barriers. This is not obvious from
comparison with critical slowing down. In particular, this
interpretation does not say why the smallest region is not a single
particle as in an Ising model, for instance. It is therefore
desirable, as a point of principle, to have simple models with a
similar phenomenology but where these issues can be addressed in a
transparent way.  Experimentally, however, it remains very difficult
to identify the relevant local rearrangements of the molecules. Even
at very low temperatures in the amorphous phase, a linear specific
heat indicates unusual elementary excitations, but was attributed to a
distribution of two-level systems for lack of more definite results
from microscopic probes. Molecular dynamics simulations, on the other
hand, have identified some collective motion of molecules along
chains,\cite{Kob} and those were argued to play a role in the glass
transition,\cite{Langer} an issue currently
discussed.\cite{Stillingerreview}

Simple models with local collective rearrangements can be defined on
lattices, for example when hard local constraints (e.g.
dimer, ice, tiling or coloring problems), together with
close-packing, prevent the motion of single particles. Typically, the
dynamics involves exchanging particles along loops of various lengths
$l$, which constitute the elementary degrees of freedom. They may be
slow (e.g. activated), and the ergodicity is not guaranteed
in general. Such models were therefore argued to be interesting
examples of glassy dynamics.\cite{Chakraborty,Castelnovo} Since they
are lattice models, they can also help to address explicitly some
issues raised by Langer,\cite{Langer} such as the ergodicity of the
dynamics when loops are involved.  For example, the ergodicity of such
models may depend explicitly on the length scales of the loops
included in the dynamics.\cite{example} Here, we employ a local
dynamics of loops and define a length scale $l^*$, such that the
dynamics of loops with length $l \geq l^*$ is able to completely
reorganize the system (i.e., decorrelates a given state) while
that with $l<l^*$ is not and generates ``static'' order. This static
order is similar to that of a spin-glass, although there is no
disorder in the model here. Contrary to spin-glass
models,\cite{BiroliBouchaud} at least when they are treated in
mean-field, the system may, however, escape its basin of energy by
passing an energy barrier of order $l^*$.  In the dimer model on the
square lattice, for instance, $l^*$ corresponds to the
\textit{shortest} loops:\cite{Hermele} Even the shortest loops
reorganize the system completely, so that the dynamics has a single
time scale. There are no growing activation barriers in this case and
the smallest existing length scale sets up the time scale.

For the degenerate three-color model on the hexagonal
lattice\cite{Baxter} (which is an ice-type model\cite{Pauling}),
however, the dynamics of the shortest loops typically leaves some
regions of space frozen.\cite{cepascanals} This happens not simply
because the shortest loops are scarce (since when a loop flips, a
neighboring configuration may in turn be flippable) but because the
frozen regions are \textit{geometrically insulated}\cite{Sriram} from
these loops. The frozen regions disappear when the next longer loops
are activated, which generates a second time scale in the
dynamics.\cite{cepascanals} We may expect a cascade of time scales,
if, instead of considering typical random states as in the degenerate
model, we consider the effect of some interactions that stabilize the
configurations with long loops.\cite{Das} Since longer loops are
slower, this leads to some interplay between dynamics and
thermodynamics. As the averaged loop length increases, does the system
generate multiple time scales?  We address this question here and show
that growing activation barriers and multiple time scales occur in the
three-color dimer model but not in the (two-color) dimer model.
This gives therefore a concrete realization of the ideas of Adam and
Gibbs in terms of a disorder-free model, but with different
interpretations.

A property of ``fragile'' glasses is precisely to have dynamics on
multiple time scales, and a relaxation time controlled by growing
activation barriers when the temperature is lowered.  Putting aside
the exact nature of the microscopic degrees of freedom, one may wonder
to what extent the present model could be a very simplified model of
``fragile'' glasses.  It has indeed three important \textit{ad hoc}
ingredients: (i) many energy minima (identified with the ``inherent
structures'' of glasses), resulting from the minimization of a strong
local interaction (modelled by a \textit{hard} constraint) that is
frustrated; (ii) a smaller interaction that lifts the degeneracy and
selects a long-loop phase with both translational and orientational
order (``crystal''); and (iii) dynamics between local energy
minima that involves local collective rearrangements of different scales,
with their time scales.  The three properties are generic and simpler
in this context than in more realistic models, where the study of the
long-time dynamics implies finding the saddle points between the
energy minima.\cite{Angelani} Here, the loop excitations have to be
seen as an oversimplified model of local collective modes, in
particular because they have a well-defined (nonrandom) sequence of
lengths, fixed by the geometry of the lattice.

\section{Length-scale-dependent static moments in energy space}
\label{DynamicsWL}

\subsection{Model and local dynamics}

We consider classical models with hard local constraints, with either
two or three state spin variables, $S_i=A,B$ (dimer model) or
$S_i=A,B,C$ (three-color model) on the bonds of the two-dimensional
hexagonal lattice (note that the bonds of the hexagonal lattice are
the sites of the kagome lattice). In the three-color model, hard
local constraints force all nearest neighbor bonds to be of different
colors,\cite{Baxter} which can be seen as a form of ice-rule. There is
an extensive number of states that satisfy the constraints and the
entropy per site was calculated exactly, $S_0=0.126375...$ in the
thermodynamic limit.\cite{Baxter} The system is paramagnetic but has
algebraic correlations because of the local constraints.\cite{Huse}

We associate an energy $E$
for a configuration that satisfies the constraints,
\begin{equation}
E = J \sum_{<<i,j>>} S_i S_j
\label{model1}
\end{equation}
where the sum runs over second nearest neighbor bonds. By convention,
the scalar product is chosen such that $S_i S_j=1$ if $S_i=S_j$ and
$-1/2$ otherwise. The three-color variables $S_i$ can be seen as
vector spins pointing along three directions at 120$^o$ (from this
point of view, the constrained states are ground states of the
frustrated kagome Heisenberg model\cite{Huse,Chandra}). A finite $J$
lifts the degeneracy of the configurations. In the following we use
$J=1$ and note $E$ the energy per site. The lowest-energy state is a
``staggered'' $Q=0$ phase ($E_0=-1$). It is a \textit{long-loop}
phase, characterized by dimers in two alternating colors along
straight infinite loops. On the other hand, the highest energy state
($E=2$), is a \textit{short-loop} phase where the spins alternate
around the hexagons (``columnar'' phase with a $\sqrt{3} \times
\sqrt{3}$ unit-cell). While both break translation symmetry, only the
former breaks orientational symmetry with a single orientation for all
the dimers of a given color.

\begin{figure}[t]
\centerline{
 \psfig{file=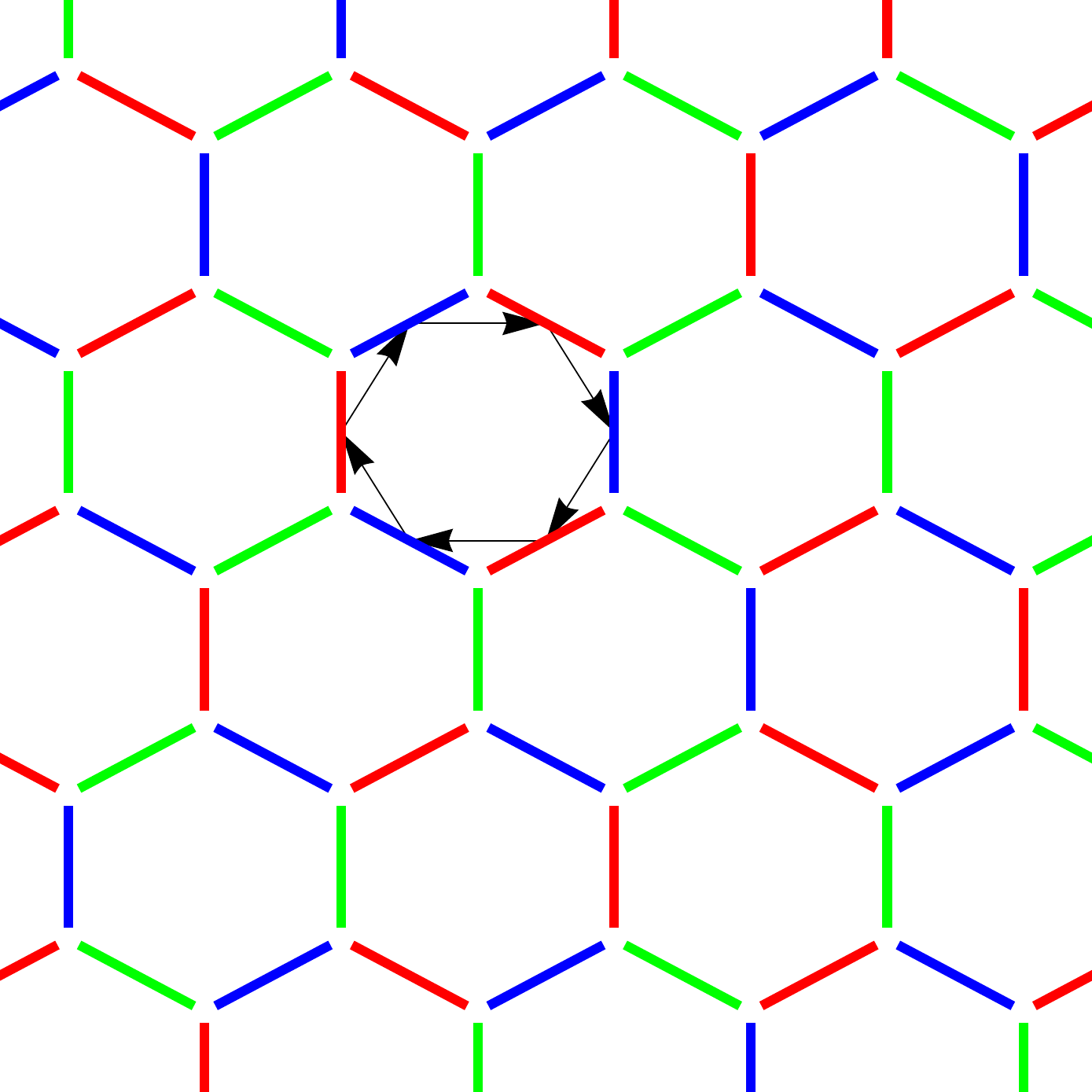,width=4.0cm,angle=-0} \hspace{.5cm}
 \psfig{file=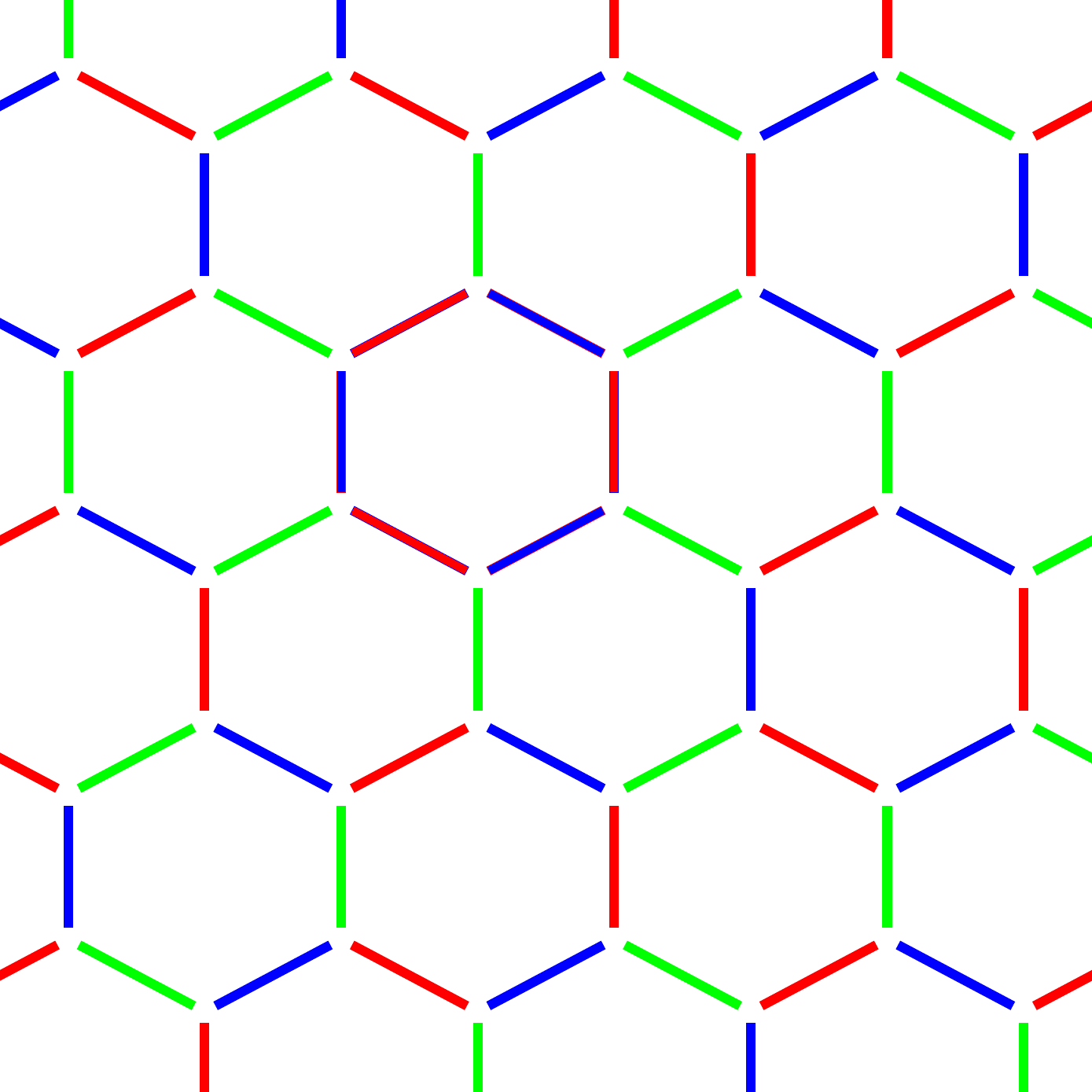,width=4.0cm,angle=-0}
}
\caption{Dynamics of local collective rearrangements (alternating
  two-color loops of size $l$) satisfying the hard local constraints (no
  nearest neighbor bond in the same state).}
\label{crr}
\end{figure}

Discrete classical models have no dynamics by themselves, but it is
interesting to consider a standard stochastic dynamics which connects
states by random local flips.\cite{effective} In the present context,
the only possible moves that are compatible with the constraints
consist of an exchange of colors along two-color closed loops (an
example is shown in Fig.~\ref{crr}). Note that there are always two
loops going through each bond, and the loop lengths are broadly
distributed.\cite{Chakraborty,Chandra} We assume that flipping a loop
is governed by an activation process with a time scale $\tau_l=\tau_0
\exp (\kappa l/T)$, which depends on the length of the loop $l$, but
we do not specify the origin of $\kappa$.\cite{cepascanals}  In fact, we further
simplify the dynamics and consider various ``levels'' of dynamics.
At each level $l$, only the loops smaller than $l$ may flip. This does not mean
that the sites belonging to longer loops never flip, because the
neighboring loops reorganize the configuration.  The dynamics here is
therefore assumed to be local, contrary to previous
works.\cite{Chakraborty,Castelnovo}

\begin{figure}[t]
\centerline{
 \psfig{file=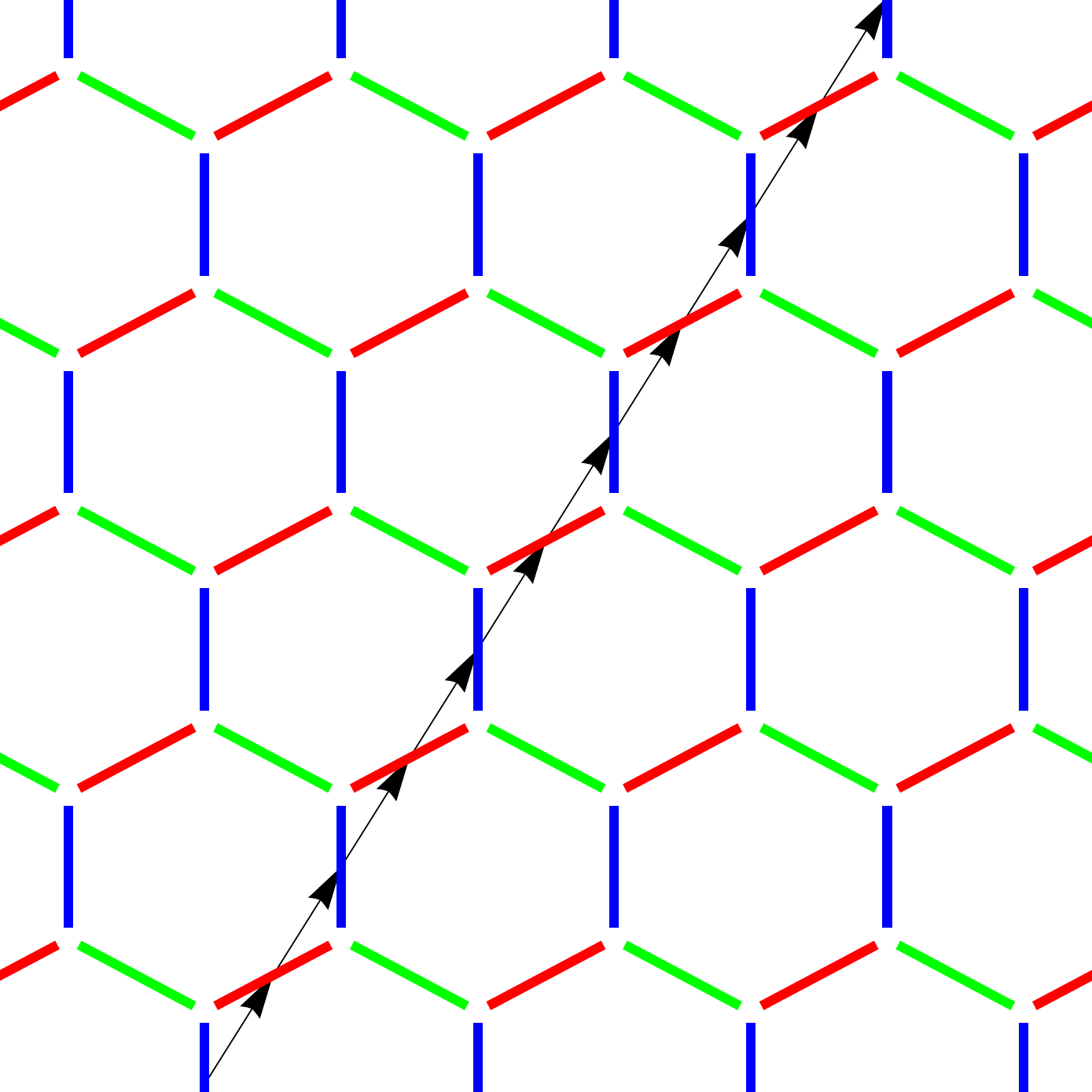,width=3.5cm,angle=-0}  \hspace{.5cm}
 \psfig{file=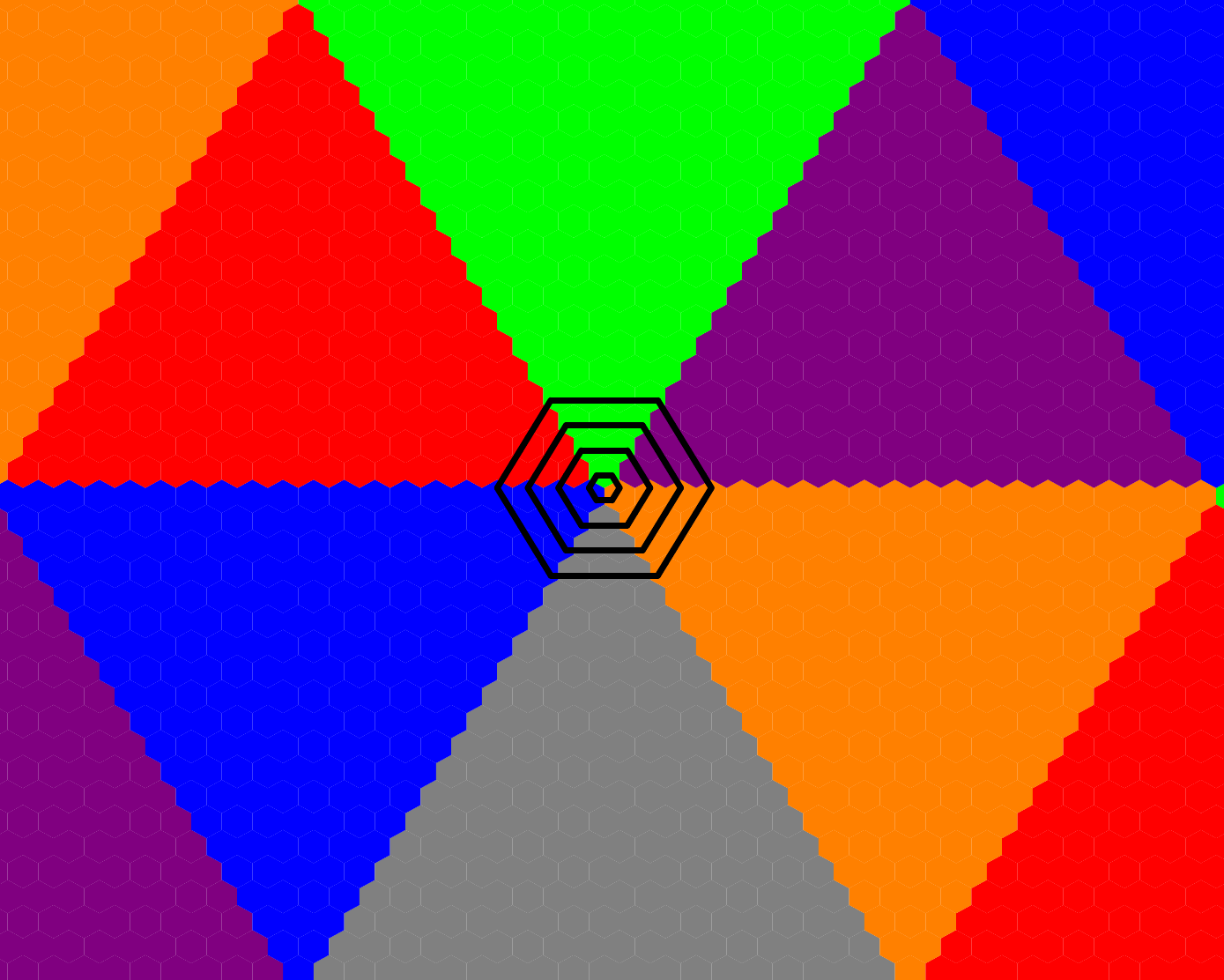,width=4.2cm,angle=-0} 
}
\caption{Right: Ground state (in the ``zero flux sector''), consisting
  of six macroscopic mono-domains related by symmetry (one is shown on
  the left), fitting in the hexagonal shape of the boundaries. While
  in a mono-domain (left), the loop degrees of freedom are straight
  and infinite, here they form concentric loops of increasing size
  (some are shown in black).}
\label{gs}
\end{figure}

We can thus address the issue of the relaxation of the dynamics by the
short-scale degrees of freedom. In particular, what is the minimal scale
that the system has to nucleate in order to relax and reach a typical
state? By including an energy for the configurations, we can study
this issue for very different states in the configuration space.  It
is indeed interesting since the short and long-loop phases are
asymmetric with respect to the length of the loops.

We start from initial states
in the microcanonical ensemble at any given energy $E$, and compute numerically the autocorrelation
function,
\begin{equation}
C_l(t) = \frac{1}{N} \sum_i \langle S_i(t).S_i(0) \rangle_E
\end{equation}
at various scales $l$, i.e., under the stochastic dynamics of
all loops up to length $l$.  The average $\langle \dots \rangle_E$ is
done over up to 500 independent initial states in each energy bin
$E$. These states are prepared by a Wang-Landau random walk
in energy space,\cite{WL} so as to cover the entire energy range. We use
finite-size clusters of hexagonal shape (that has the same symmetries
as the infinite lattice) with periodic boundary conditions, and sizes
up to $L=30$ for the dynamics ($N=9 \times L^2$) and $L=26$ for the
computation of the density of states. For larger sizes, the density of
states does not converge in the entire energy range (with up to $5
\times 10^6$ Monte Carlo sweeps [MCS] for each Wang-Landau iteration)
and we have used smaller energy intervals by starting from special
states. When $l$ is large enough, the autocorrelation relaxes to zero from $C_l(0)=1$, in a few MCS. For
small $l$, we use 200 MCS to allow for convergence
(except for $l=6$ and high-energy states $E \sim 2$, where the
convergence is much slower and we use $4 \times 10^3$ MCS).

Since the dynamics is local, we reject in particular the infinite
loops that wind around the periodic boundary conditions. As a
consequence, the dynamics conserves some topological
numbers,\cite{Castelnovotopo} and we restrict the study to the ``zero-flux
sector''. This is without loss of generality since the lowest energy
state in the zero-flux sector is not the mono-domain $Q=0$ state
mentioned above (which is in the highest-flux sector) but consists of
six triangular mono-domains, arranged in such a way as to be
compatible with the hexagonal shape (see Fig.~\ref{gs}). Its energy
per site $E$ is higher by a ``surface'' energy term, $E-E_0=3/L$, that
vanishes in the thermodynamic limit. It is interesting to consider the
latter since it allows to address the question as to whether it can be
connected from a typical state by a \textit{local} dynamics, whereas
the former cannot.

\begin{figure}[t]
\centerline{
 \psfig{file=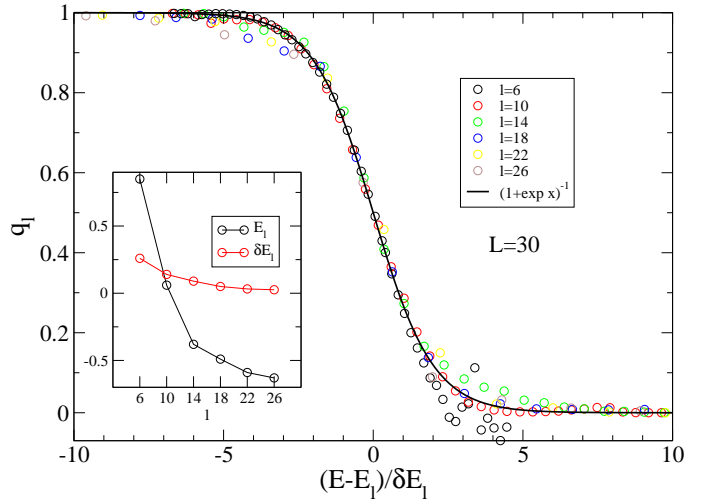,width=9.5cm,angle=-0}
}
\caption{Static moments, $q_l=\lim_{t \rightarrow \infty} C_l(t)$ for the dynamics including loops up to length $l<l^*$, as functions of rescaled energy  $x=(E-E_l)/\delta E_l$, with parameters given in the inset. They  collapse approximately onto a single curve (solid line).}
\label{q}
\end{figure}

For typical random states of dimer models, the dynamics is generally
expected to be diffusive with $C_l(t) \sim t^{-\alpha}$ at long
times.\cite{Henley} For the degenerate three-coloring model, this is
indeed the case when $l$ is large enough, with an exponent
$\alpha=2/3$.\cite{cepascanals} For smaller values of $l$, we now show
that the diffusion is blocked and the autocorrelation saturates to a
finite value after some short time,
\begin{equation}
q_l=\lim_{t \rightarrow \infty} C_l(t).
\end{equation}
A finite value of $q_l$ indicates an overlap between the initial state
and the final state, i.e., a static frozen moment, as in
spin-glasses. $q_l$ is given in Fig.~\ref{q}, for different $l$ and as
a function of the energy $E$ of the initial state. They increase
continuously below energy thresholds that depend on $l$, and reach
values close to one at lower energies.  The figure shows that the
curves collapse approximately onto a single function $f(x) \approx
(1+\exp x)^{-1}$ of the rescaled parameter $x=(E-E_l)/\delta E_l$
(solid line in Fig.~\ref{q}), where $E_l$ and $\delta E_l$ are given
in the inset. Note that for $l=6$ at high-energy, there are some
fluctuations in the points resulting from low statistics in this
energy range.  The collapse shows some self-similarity
of the dynamics at different scales.

\subsection{Growing barriers and multiple time scales}

We define an ergodic or relaxation length scale $l^*$ such that
$q_{l^*}=0$, but $q_{l} \neq 0$ for $l<l^*$.
Since the loops of length $l^*$ have to flip before the system can
fully relax, the relaxation time of the system is given by
$\tau_{l^*}$.  In other words, when $q_l\neq 0$, the system is trapped
into some basin of energy, and moves within smaller basins under the
dynamics of loops of length $l<l^*$. It is only when the system passes
energy barriers of order $l^*$ that it escapes its basin into a
liquidlike regime.

\begin{figure}[t]
\centerline{
 \psfig{file=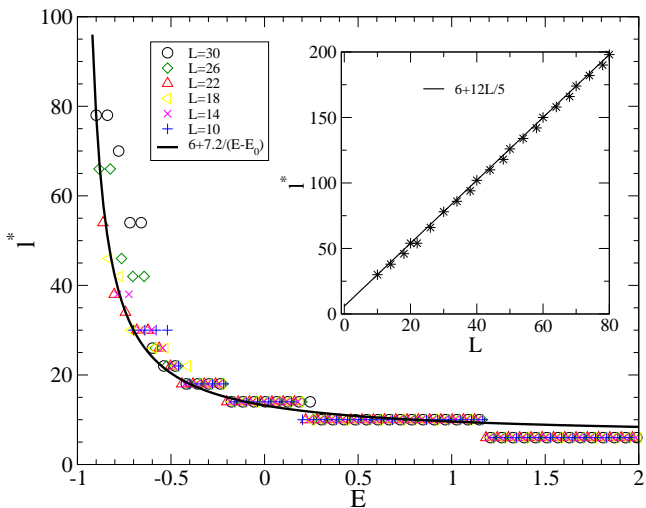,width=9.cm,angle=-0}
}
\caption{Minimal length $l^*$ of loops that need to be included in the dynamics to decorrelate a state with initial energy $E$: $l^*$ grows at low energies generating an additional time scale at each step.
Inset: Diverging $l^* \sim L$ for the ground state (the solid line is exact when $L$ is a multiple of ten).}
\label{lstar}
\end{figure}

\begin{figure}[t]
\centerline{
 \psfig{file=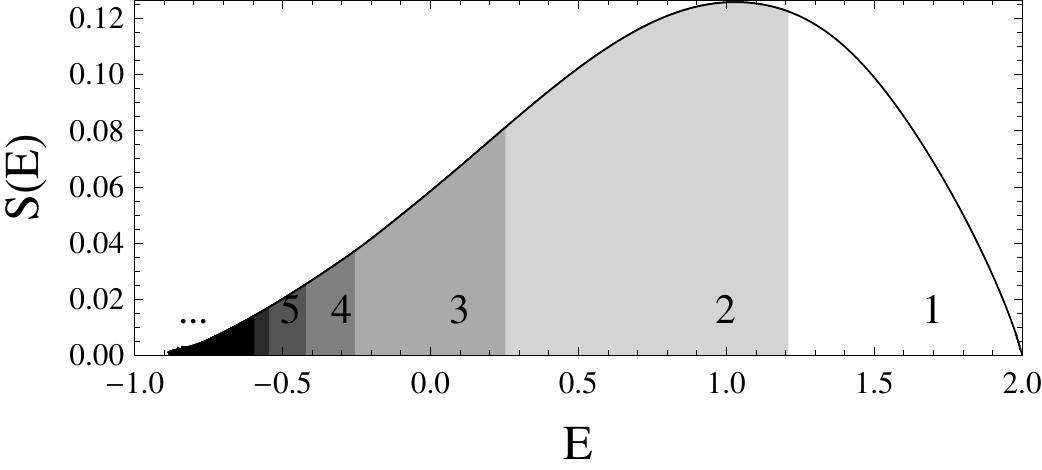,width=8.5cm,angle=-0} }
\centerline{
 \psfig{file=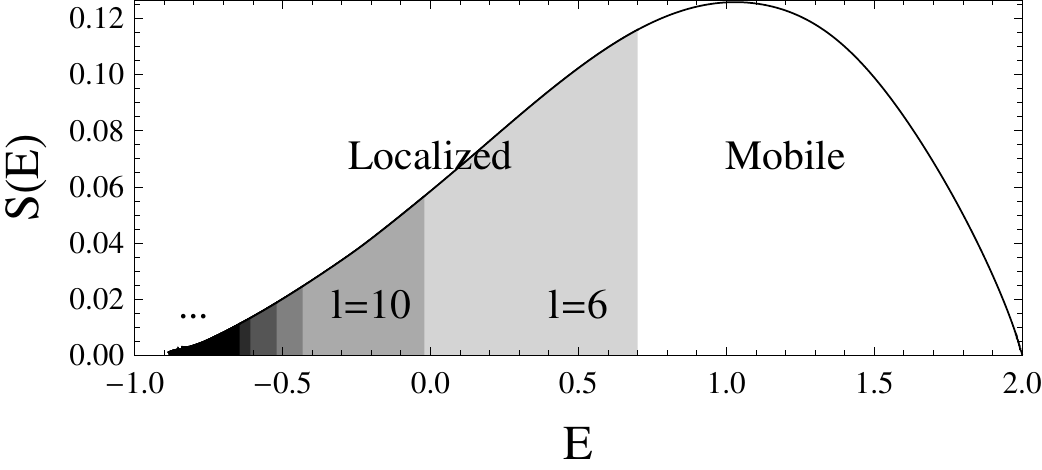,width=8.5cm,angle=-0} }
\caption{Logarithm of the density of states, $S(E)$, numerically determined by the Wang-Landau algorithm for $L=26$.  The number of time scales in the dynamics and their energy thresholds are indicated (top panel), as well as the (classical) ``mobility edges'' for loops of length $l$  (bottom panel). The maximum of $S(E)$ is 0.1258, very close to Baxter's result, 0.1264, in the thermodynamic limit.}
\label{lng}
\end{figure}

We have computed $l^*$ for the three-color model as a function of
the energy of the initial state (Fig.~\ref{lstar}), by using the data
of Fig.~\ref{q}.  In practice, the thresholds slightly depend on the
definition of $q_l \neq 0$ (which we have taken to be $q_l>0.2$).  It
is a discontinuous function since the allowed values of $l$ are
a multiple of four on this lattice, starting from the smallest hexagonal
loops with $l=l_1=6$.  For the typical random states at $E \approx
1.03$ (corresponding to the maximum of entropy, see Fig.~\ref{lng}),
we have $l^*=10$, giving two time scales, as shown
earlier.\cite{cepascanals} By reducing the energy further, $l^*$
increases to very large values. This corresponds to growing activation
barriers in the relaxation time, $\tau_{l^*}$.  Furthermore, the
motion of loops with $l<l^*$ reorganizes the system partially (with
the values of $q_l$ given in Fig.~\ref{q}), on time scale,
$\tau_l$. The dynamics is, therefore, characterized by multiple
intermediate time scales, the number of which increases with $l^*$.
The energy landscape becomes more and more hierarchical at low energy
with basins within basins.

For the ground state, $l^*$ is growing with $L$ (inset of
Fig.~\ref{lstar}), in particular we find that $l^*=l_1(1+2L/5)$ (solid
line) is exact when $L$ is a multiple of ten, which we explain
below. It is therefore diverging in the thermodynamic limit and the
ground state cannot be melt in a typical state by local processes. The
ground state is truly metastable in this model.  For intermediate
energies, we find that
\begin{equation}
l^*=l_1+ \frac{\beta}{E-E_0}
\label{lstarE}
\end{equation}
with $\beta=7.2$, is a rather good approximation in the entire energy
range, bare the discreteness of the loop lengths (solid line in
Fig.~\ref{lstar}). We show that it is exact for all the
finite-size ground states.  From the numerics, it is not clear whether
$l^*$ could diverge at an energy $E>E_0$.  The statistics in averaging
$C_l(t)$ is low for $E<-0.5$ because it is more difficult to generate
independent states in this energy range.  We construct in Sec.
\ref{excited} some excited states with $E>E_0$, where $l^*$ is also
infinite but they have a finite overlap with the ground state.
Nevertheless, Fig.~\ref{lstar} provides a connection between the
energy of the system and its dynamics.

The behavior of $l^*$ can be simply understood for the lowest and
highest energy states. First, the state at $E=2$ has \textit{all} its
hexagons flippable: It is therefore clear that $l^*=6$. Incidentally,
the averaged loop length is also $\langle l \rangle=6$. For the ground
state, on the other hand, the two-color loops are straight lines in
each mono-domain and form concentric structures around the three
special points where all the domains join (see Fig.~\ref{gs}). Their
lengths increase as $l_k=l_1(2k-1)$ with $k=1,\dots,L$ (in particular,
the averaged loop length $\langle l \rangle = L l_1$ and the size of
each domain are diverging in the thermodynamic limit). However the
concentric loops of small sizes create only localized rearrangements
in small zones and create no new flippable loop of the same length. It
is only when the loop size is of the order of the distance between two
such zones that a complete reorganization becomes possible. Since the
domains are macroscopic, we therefore conclude that $l^* \sim
L$. Careful examination of the state shows that $l^*=l_1(1+2L/5)$ when
$L$ is a multiple of ten, as found numerically. Since the ground state
has energy $E=E_0+3/L$, we may rewrite $l^*=l_1+\beta/(E-E_0)$ with
$\beta=6l_1/5=7.2$, which is precisely Eq.~(\ref{lstarE}). This formula
is therefore exact for all the finite-size ground states (when $L$ is
a multiple of ten), and is a good approximation in the entire energy
range.

\begin{figure}[t]
\centerline{
 \psfig{file=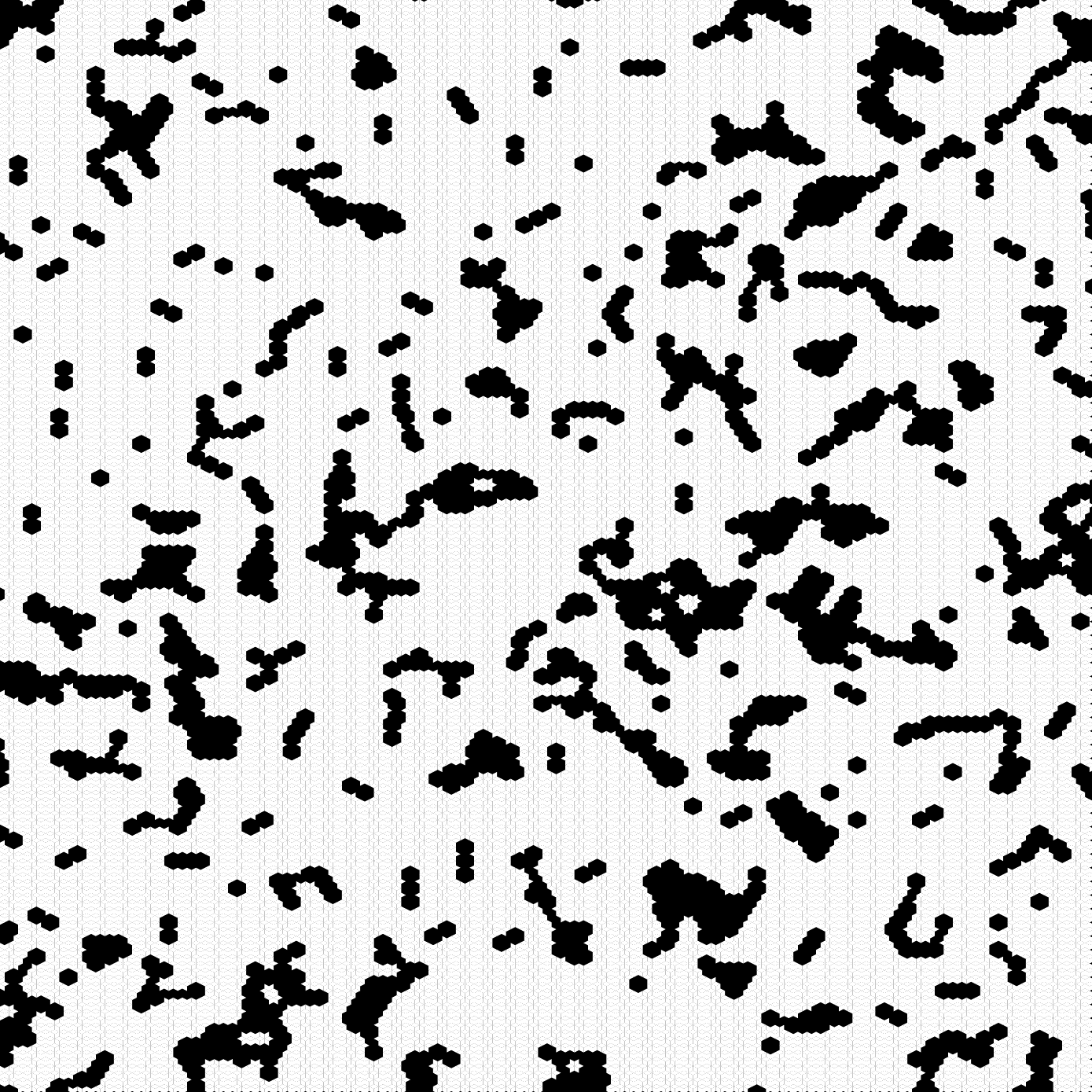,width=4.3cm,angle=-0}
 \psfig{file=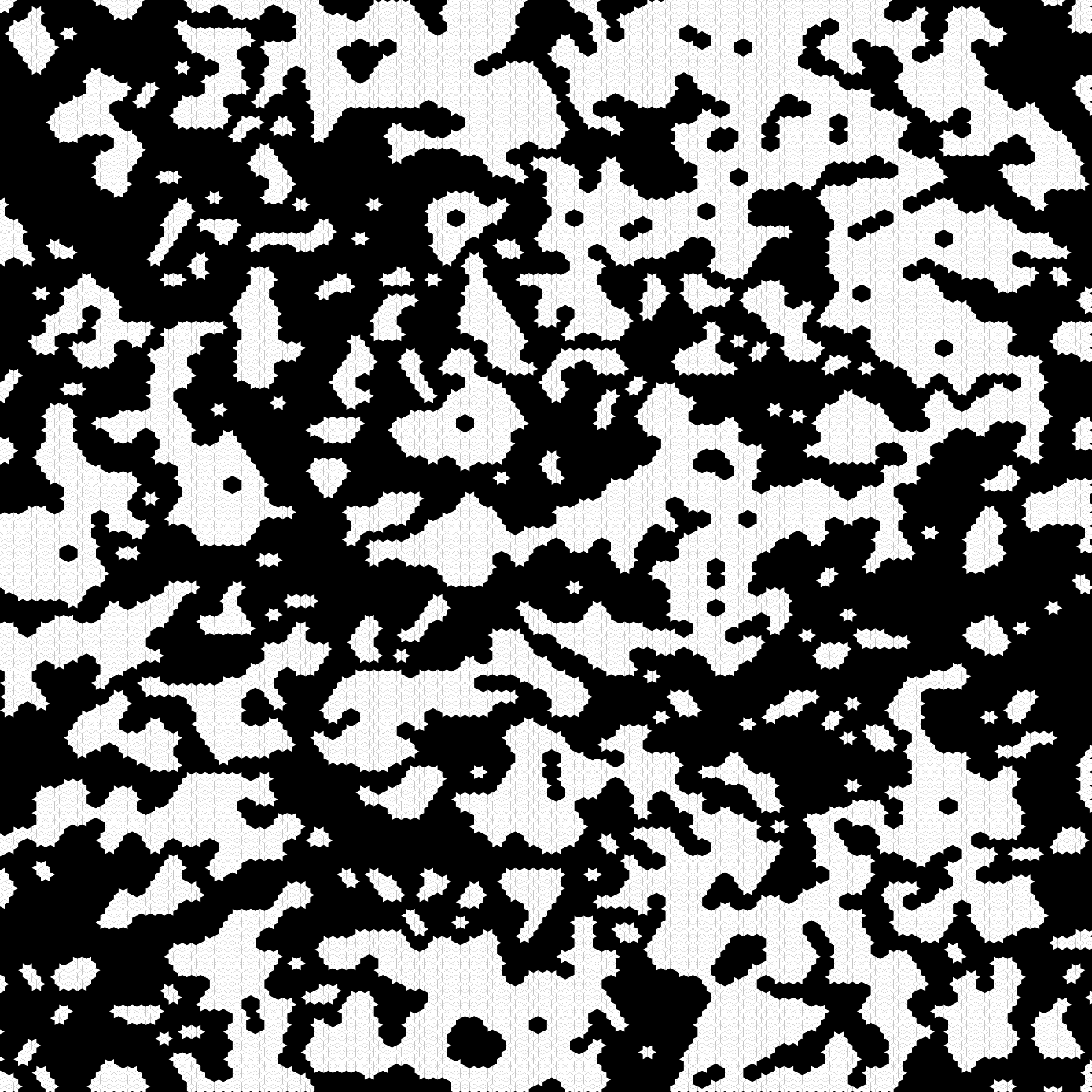,width=4.3cm,angle=-0}
}
\centerline{
 \psfig{file=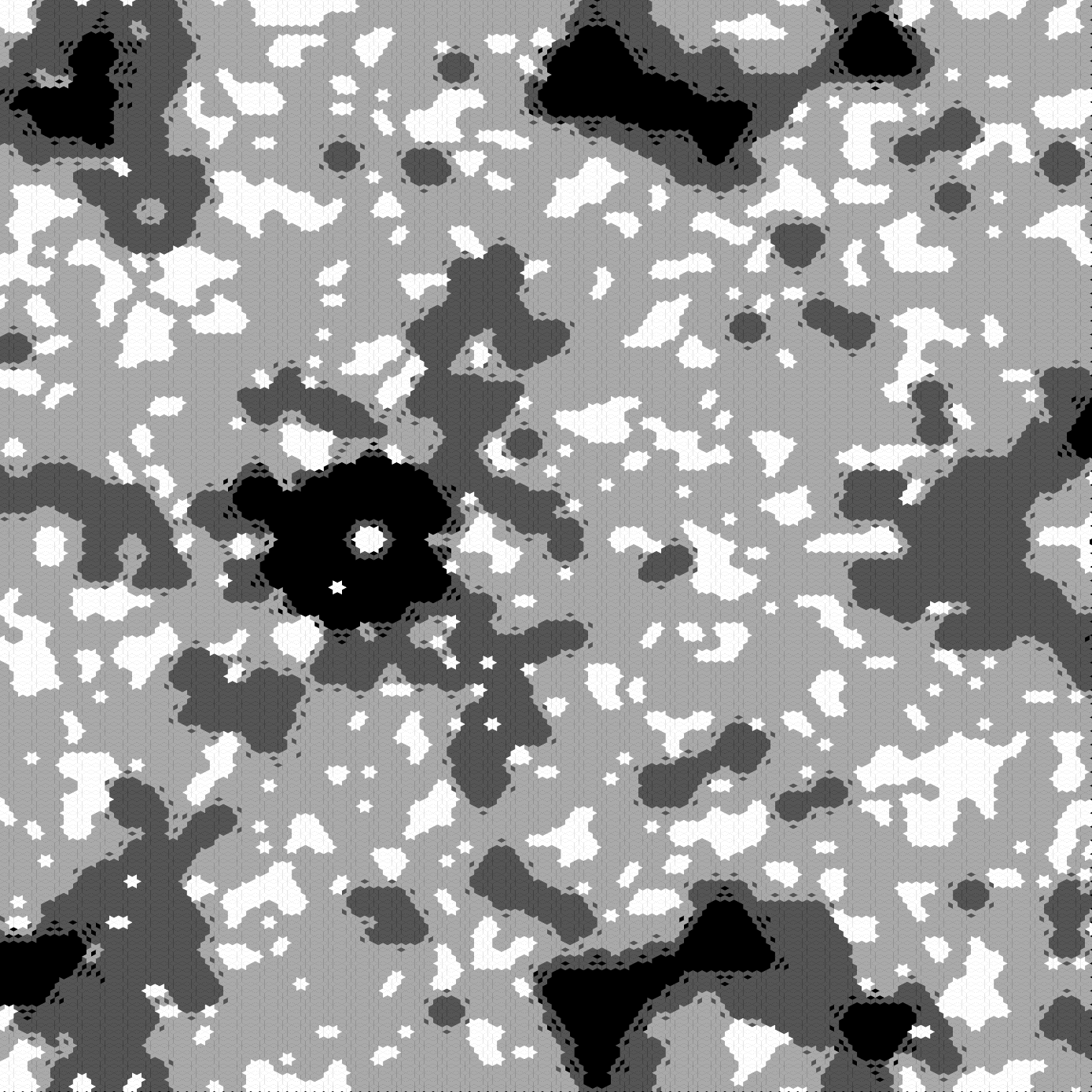,width=4.3cm,angle=-0}
 \psfig{file=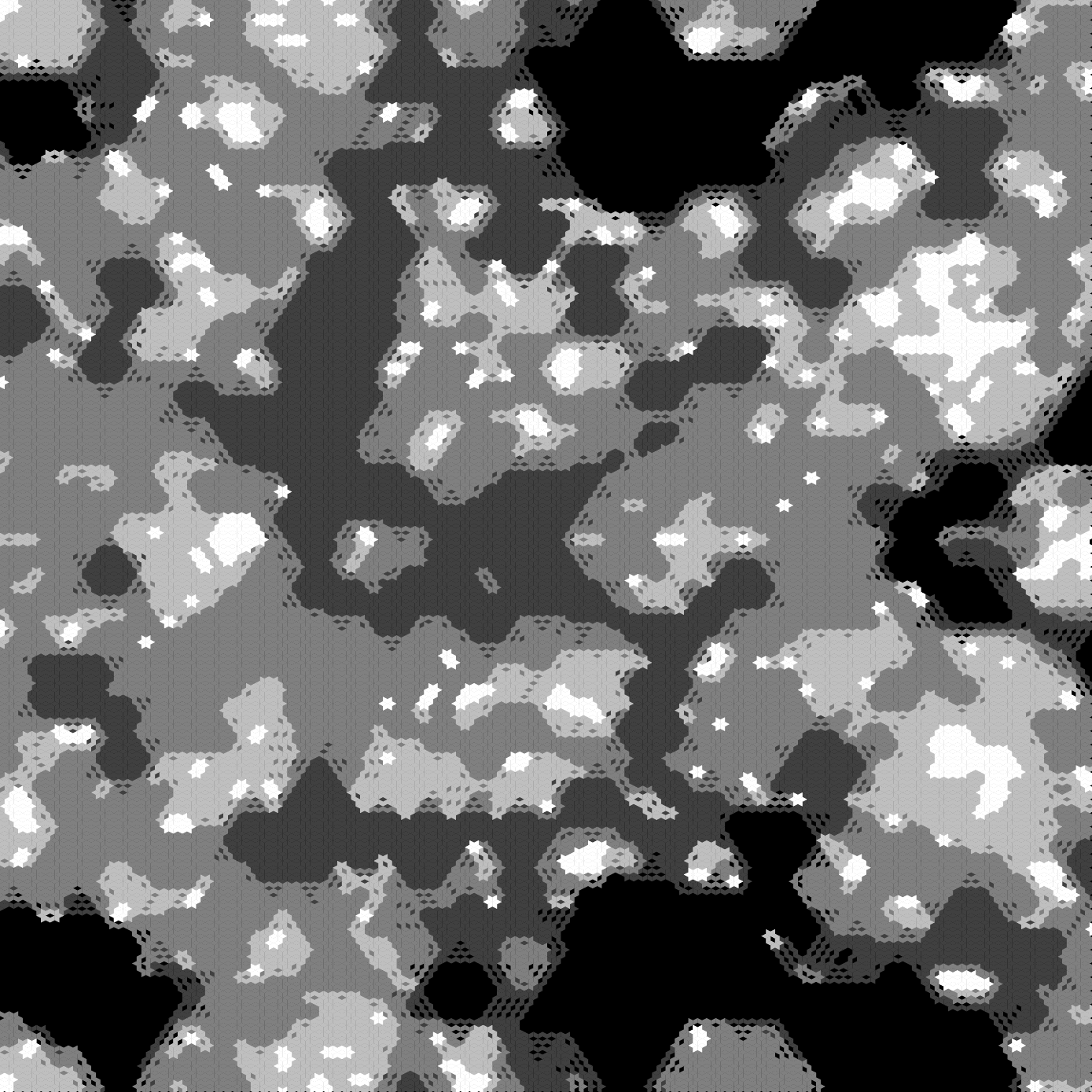,width=4.3cm,angle=-0}
}
\centerline{
 \psfig{file=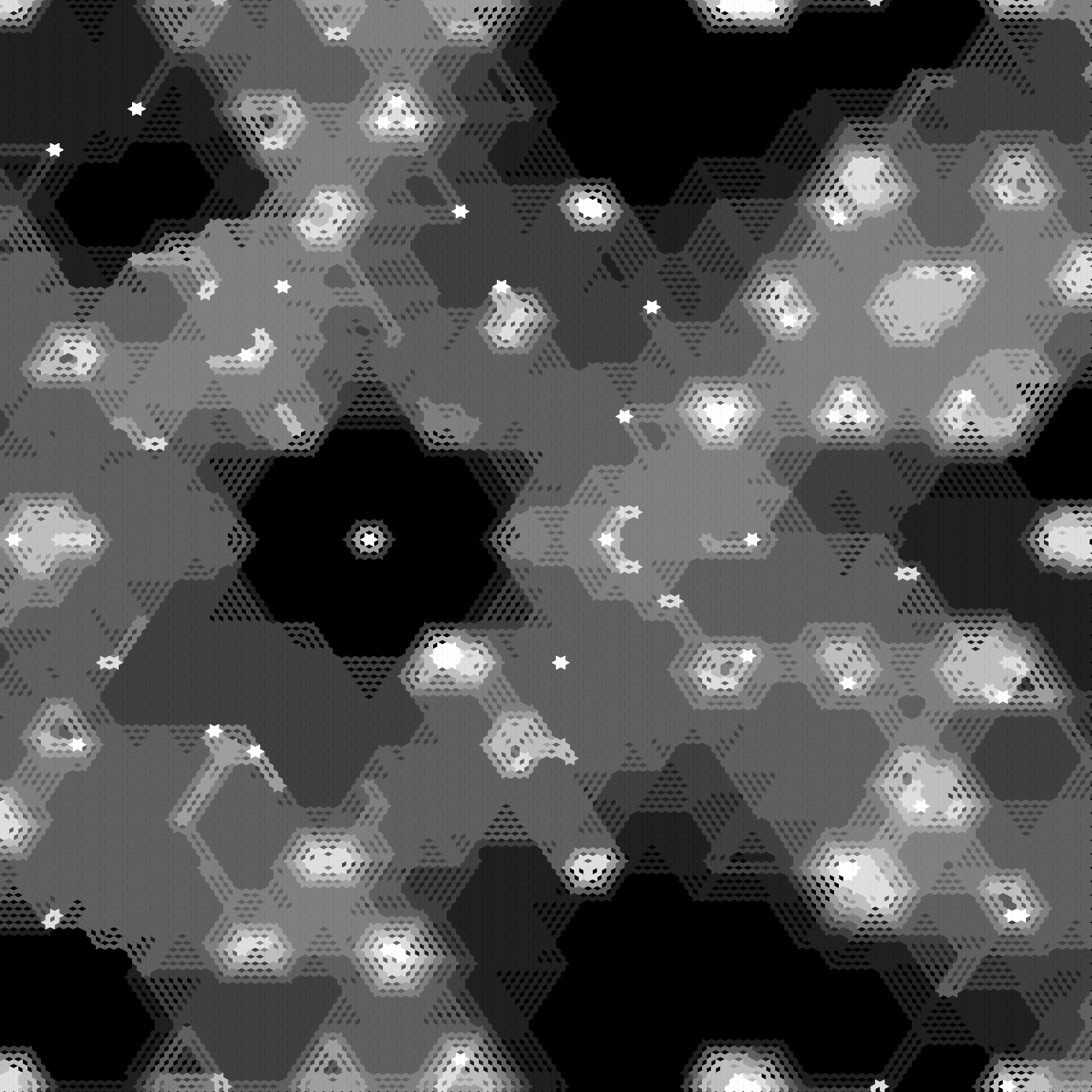,width=4.3cm,angle=-0}
 \psfig{file=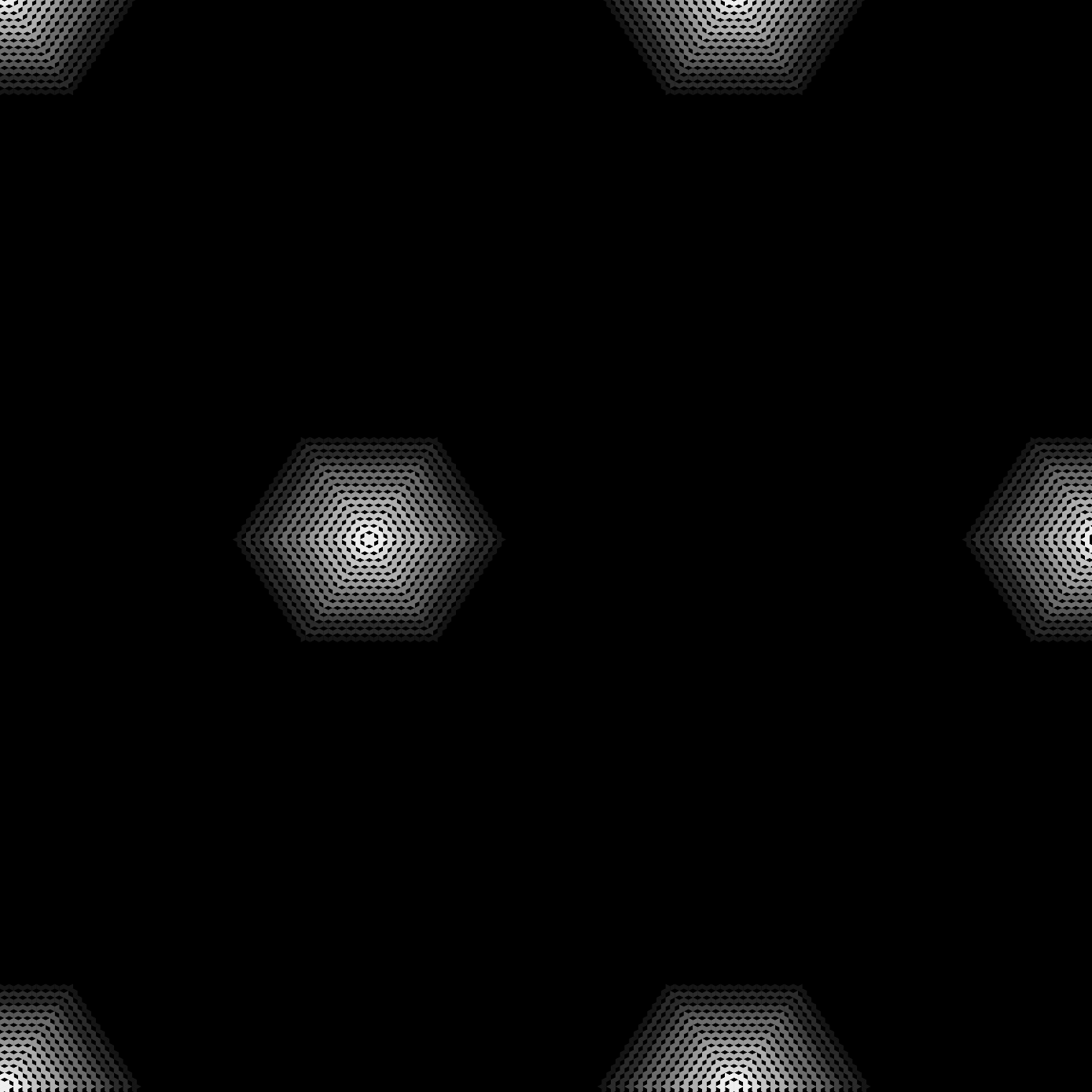,width=4.3cm,angle=-0}
}
\caption{Hierarchies of frozen regions developing from high energy (typical state, top left) to low energy (ground state, bottom right). Each deeper level of gray corresponds to sites that move on a longer time scale $\tau_l$ ($l<l^*$), i.e., a region insulated from smaller faster loops ($<l$). There are energy thresholds $E_l$ for the appearance of time scales and thresholds for the localization (percolation) of the degrees of freedom of length $l$ in isolated pockets (see Fig.~\ref{lng}).}
\label{map}
\end{figure}
 \begin{figure}[t]
\centerline{
 \psfig{file=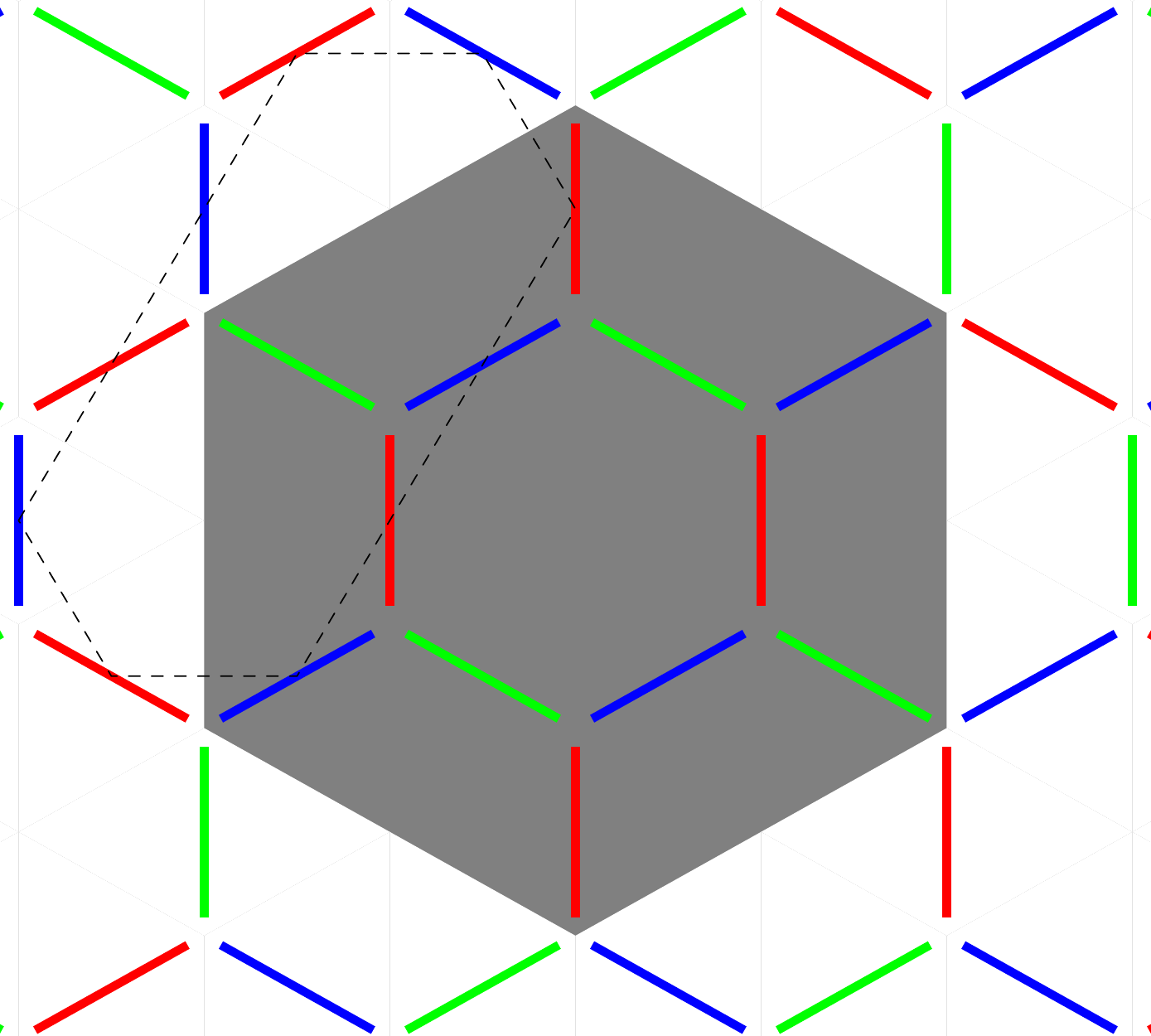,width=4cm,angle=-0} \hspace{.5cm}
 \psfig{file=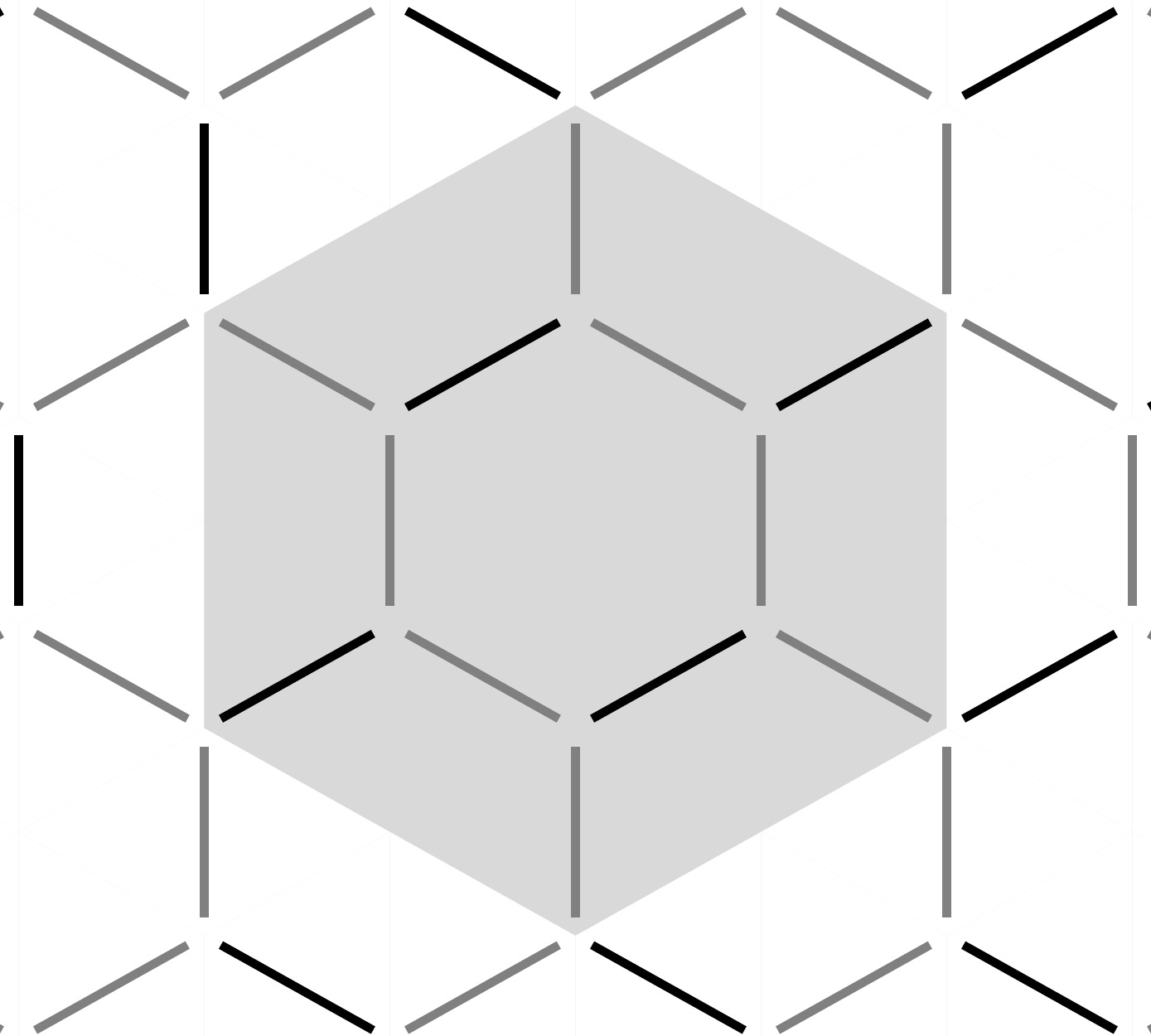,width=4cm,angle=-0} }
\vspace{.5cm}
\centerline{  
\psfig{file=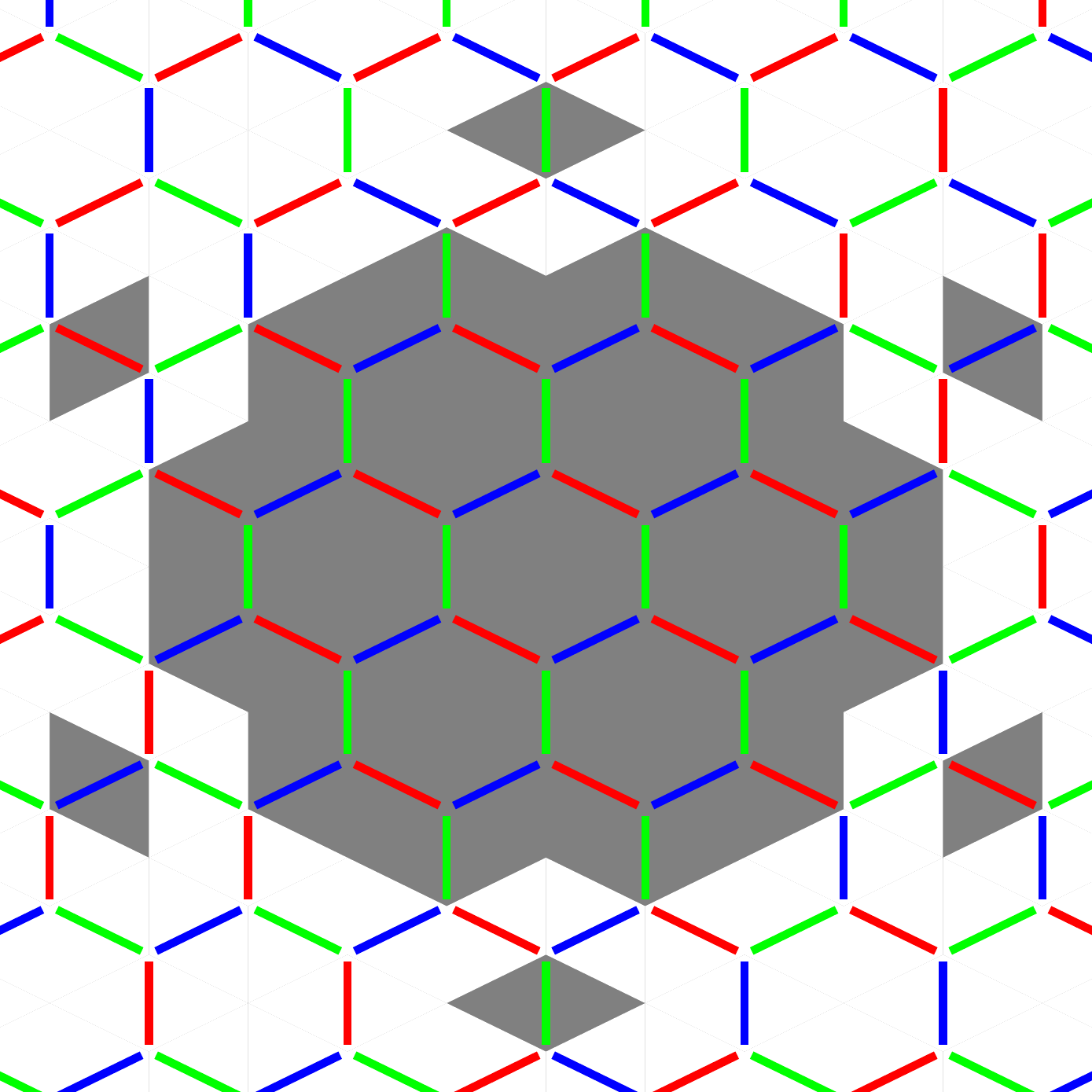,width=4cm,angle=-0} \hspace{.5cm}
\psfig{file=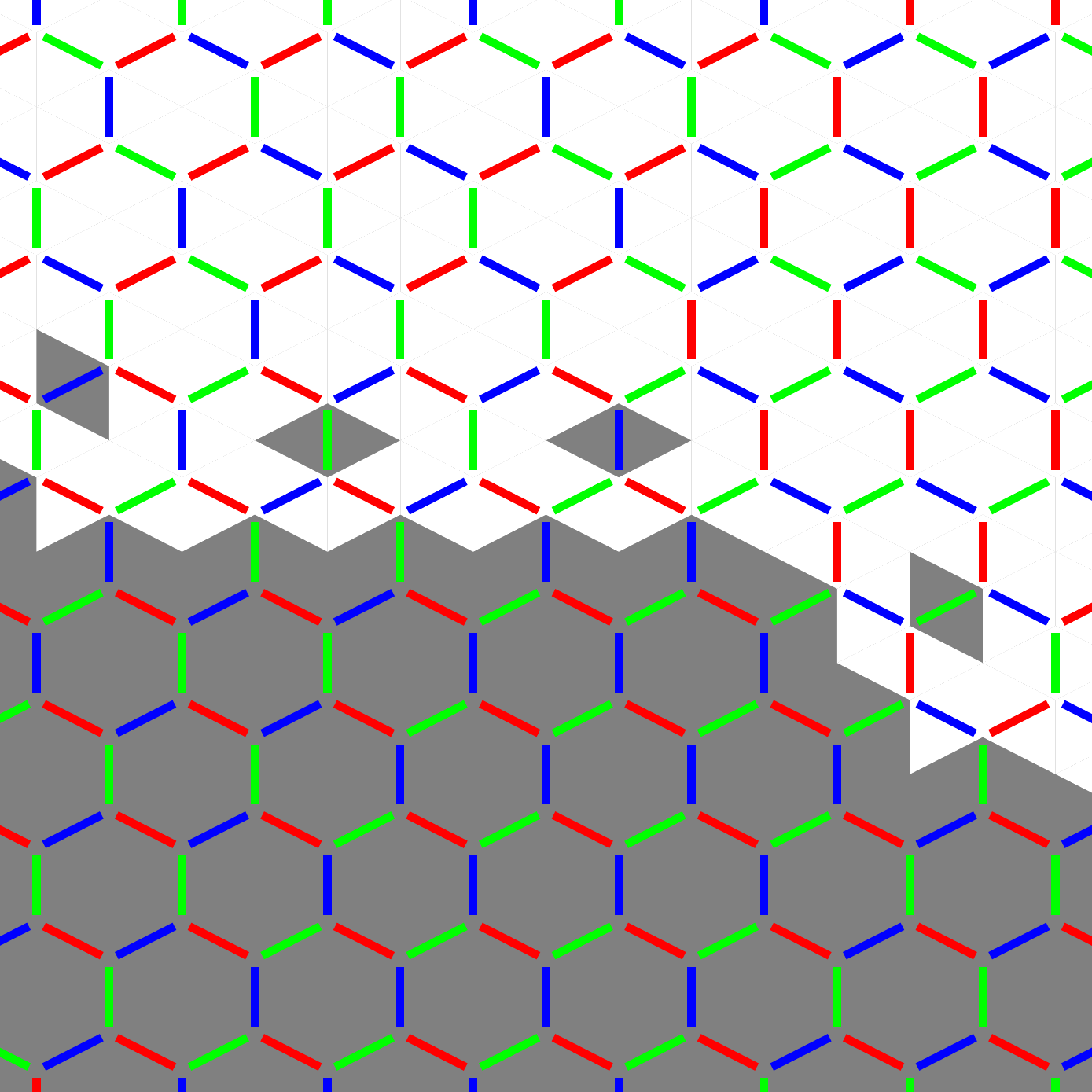,width=4cm,angle=-0} }
\caption{Simple examples of regions insulated from $l=6$ (top left) or $l=10$ (bottom left) degrees of freedom. In the dimer model, there is no insulated cluster (top right). Insulated regions are not necessarily a perfect domain of the long-loop phase (see, for instance, the bottom right figure).}
\label{jammed}
\end{figure}

We may now ask whether Eq.~(\ref{lstarE}) is generic for models with long-loop ground states. When the averaged loop length $\langle
l \rangle$ and the size of the ordered domains diverge at low
energy, does the system necessarily have a diverging $l^*$? 
In order to understand this point, we consider the
equivalent dimer model on the hexagonal lattice. It has a similar
long-loop phase (staggered phase), which, in the zero-flux sector, has
three domains made of the three long-range dimer orientations.  By
contrast, in this case, we find that $l^*=l_1$. This happens because
flipping the single flippable hexagon located where the three domains
meet generates in turn neighboring flippable hexagons which propagate
in the entire system. This is very different from the three-color
model studied above where flipping the same loop does not generate
flippable neighboring loops. The important point is therefore not the
density of small loops (a single loop may be sufficient) nor the
presence of large domains with a diverging $\langle l \rangle$ but
whether these loops are able to reorganize the system on a
global scale. We now explain why they cannot in the three-color
model.

\subsection{Insulated regions}

We study how the static moments at the different levels of the
dynamics are structured in real space, in frozen regions. In order to
do this, we record the spins which do not move within the time of the
simulation (the persistent field) on different scales. In
Fig.~\ref{map}, each figure shows a real-space map of the persistent
field at different levels $l<l^*$ of the dynamics (each level is shown
in gray scale) for the same initial state.  The figures correspond to
different initial states with different energies which range from a
typical state (top left) down to the ground state (bottom right). For
the typical state, the two time scales are visible, with the fast
$l=6$ (in white) and slow $l=10$ (in black) regions. For lower energy
states, we see a developing hierarchy of regions, with higher-level
regions fully insulated from all shorter loops. Each region at level
$l_0$ is inaccessible to the loops of length $l<l_0$, but not to loops
of length $l_0$. They are therefore frozen on all time scales, $\tau_l$
($l<l_0$), but have some dynamics on the longer time scale
$\tau_{l_0}$.  The origin of this behavior is easy to understand by
looking at the configurations.

A region is insulated, when its border is protected from the flippable
degrees of freedom, whatever configuration the outside may take in
the course of the dynamics. Examples are given in
Fig.~\ref{jammed}. The simplest example (top left) is a 12-site
cluster of sites insulated from the loops of length 6: The central
hexagon is not flippable (it is not a two-color loop) and the
neighboring hexagons will never be flippable, since they have already three
colors on the border. We therefore immediately see that this is not
the case for the equivalent (two-color) dimer model (top right):
There is nothing that prevents the border from being flippable.  The
insulated cluster, drawn on top left, is not insulated from loops of
length 10 (one is shown by a dashed line). The simplest cluster
insulated from loops of length 10 is shown at the bottom left. It has
48 sites and there is no loop of length 6 nor 10 inside or
on its border. The smallest flippable one is 14. The simplest
cluster insulated from loops of length 14 has one 108 sites, etc.  Although all these clusters have the local order of
the ground state with long loops, larger insulated regions do not
(see an entire region insulated to loops 10 in Fig.~\ref{jammed},
bottom right): The insulated regions do not coincide
with the domains of the ordered state.

\vspace{-1cm}
\subsection{Percolation of the insulated regions}

When the static moment $q_l$ becomes large, the corresponding frozen regions also become large and percolate. For example, at $E=0.6$, the frozen regions form an infinite percolating cluster (top right in Fig.~\ref{map}). We find indeed that the first percolation threshold occurs at $E \sim 0.7$, where $q_6=0.64 \equiv q_p$. The density of frozen sites is $0.44$, smaller than the standard bond-percolation threshold, $0.6527$, on the hexagonal lattice (or site-percolation threshold of the kagome lattice), but there are correlations here and the threshold is a nonuniversal quantity.
As a consequence, the $l=6$ degrees of freedom are now localized in some pockets in space, and become more and more localized  as the energy further decreases (see Fig.~\ref{map}). As the energy is lowered, the longer scales also get successively localized and we have a cascade of percolation transitions.  We have not computed the precise (classical) ``mobility edges'' for all scales, and some schematic ones are shown in Fig.~\ref{lng}, simply obtained by using $q_l = q_p$. For the ground state (bottom right), there is an infinite number of scales up to $l=l^* \sim L$, that are localized in the zones shown. Because of this localization, no local scale can rearrange the system globally. 

This model therefore provides an example where classical dynamical
degrees of freedom get localized in the absence of disorder. This implies,
in particular, that the energy injected in a given state cannot diffuse
on the time scale $\tau_l$ unless the corresponding region at level
$l$ percolates. 

\section{Thermal equilibrium and metastability}

Among the states of energy $E$ studied above, we can study the most probable states at thermal equilibrium at a temperature $T$. We
show below that the system undergoes a first-order phase transition at
thermal equilibrium, between a paramagnetic phase and the ground
state represented in Fig.~\ref{gs}. However, since we found that the ground state is dynamically unreachable by
the local processes we have considered (the system has to pass a macroscopic barrier of order $\sim L$), the ergodicity
is broken below $T_m$. Depending on thermal preparations, any energy
state of the system may be relevant.

\subsection{First-order transition at equilibrium}
\label{fopt}

We have computed the equilibrium free-energy $F(E,T)=E-TS(E)$ from
$S(E)$ shown in Fig.~\ref{lng}. The logarithm of the density of state
$S(E)$ is obtained by the Wang-Landau Monte Carlo algorithm.\cite{WL}
In these simulations, we have flipped the loops irrespective of their
sizes, according to the acceptation ratio that makes it possible to reach flat
histograms in energy space. We have used a square root reduction of
the modification factor at each Wang-Landau step and checked the
convergence of $S(E)$. We have obtained the density of states in the
entire energy range for $L \leq 26$, but not for larger sizes. We also mention that we could not directly equilibrate
the system at low temperatures with the standard loop Metropolis algorithm for $L>20$
(with 10$^6$ MCS per temperature). This is, in fact, similar to previous works
where the equilibration of long-loop phases in the triangular Ising
model\cite{Dasgupta} or the three-color
model\cite{Chakraborty,Castelnovo} is not possible for large sizes,
even with a nonlocal dynamics.

As a test of the results, we note that the maximum of $S(E)$
obtained for $L=26$ is at $0.1258$, within 0.5\% of Baxter's
thermodynamic result, $S_0=0.126375$... Simulations in the restricted
energy range $[0,2]$ up to $L=40$ leads to a finite-size extrapolation
of $0.12628$ or $0.12647$, depending on whether we include a
$1/L$ contribution in the fit.

\begin{figure}[t]
\centerline{
 \psfig{file=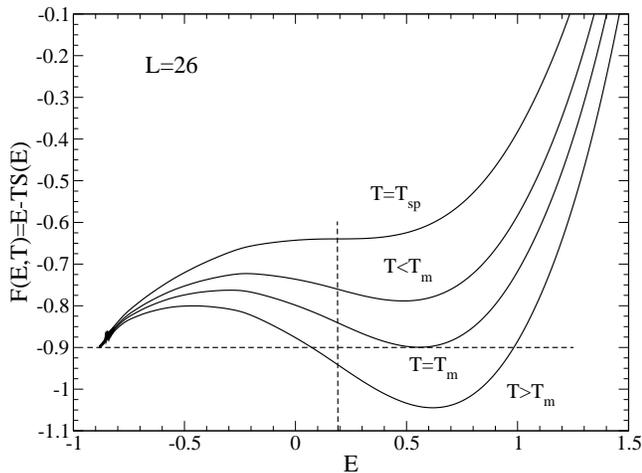,width=8.5cm,angle=-0} }
\caption{First-order phase transition at thermal equilibrium, obtained from Wang-Landau simulations. There are two minima in the free-energy $F(E,T)=E-T S(E)$ down to $T_{sp}=11.0$ (upper curve; vertical dashed line: $E_{sp}=0.19$), and equal at $T=T_m=13.6$ ($L=26$).}
\label{freeenergy}
\end{figure}
\begin{figure}[t]
\centerline{
 \psfig{file=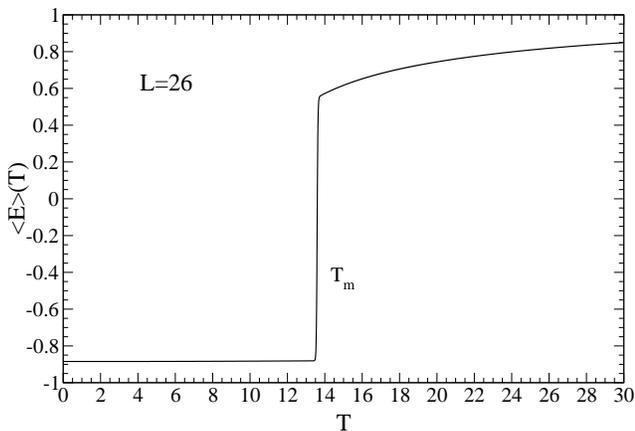,width=8.5cm,angle=-0} }
\caption{Jump in the internal energy as a function of temperature at the first-order transition, at equilibrium, from Wang-Landau simulations.}
\label{energyjump}
\end{figure}

In Fig.~\ref{freeenergy}, we show that the free-energy has two minima
above a spinodal temperature $T_{sp}=11.0$, where the paramagnetic
phase becomes unstable. The two minima have the same free-energy at
$T=T_m(L=26)=13.6$. Finite-size scaling leads to $T_m=14.50$ in the
thermodynamic limit. This is the predicted melting temperature where
the internal energy of the system $\langle E \rangle(T)$ jumps (see
Fig.~\ref{energyjump}, computed from the density of states by using canonical
formulas\cite{WL}). The ground state energy is indeed $E=E_0+3/L$ as
explained above.  The simple argument for the first-order transition
is that of an attraction between the long loops: Fitting two long
loops as nearest neighbors costs twice the boundary energy, whereas
fitting them apart costs four times the boundary energy.

\subsection{Metastability and absence of critical droplets}
\label{meta}

The first-order transition takes place, provided that thermal
equilibrium can be achieved.  As the temperature is lowered below
$T_m$, the system remains metastable in the supercooled phase down to
$T_{sp}$. Its relaxation time is controlled by $\tau_{l^*}$, which
increases faster than Arrhenius.  At $T_{sp}$, we have $\langle E_{sp}
\rangle=0.19$ (vertical dashed line in Fig.~\ref{freeenergy}), so that
we obtain $l^*=14$ from Fig.~\ref{lstar}.  The enhancement of the
energy barrier is only a factor of $14/6$, when the liquid looses its
stability.  Usually, in a first-order transition, a critical droplet
nucleates and grows. Here, this does not take place because the growth
of droplets (as evidenced from the growth of $\langle l \rangle$) is
accompanied by the growth of $l^*$, i.e., a slowing down of
the dynamics. There is no critical size of the droplet above which the
growth will be fast. For $T<T_{sp}$, the dynamics is
out of equilibrium, and the system evolves in the hierarchical energy
landscape that we have described, without being able to reach its ground state.

\subsection{Linear specific heat due to localized rearrangements}
\label{excited}

We argue that the localized elementary excitations may give rise to a  specific heat that is linear in temperature, in the metastable low-energy states.
For this, we first consider the ground state (Fig.~\ref{gs}) and its excited configurations, obtained by flipping loops with $l<l^*$, so that the ground state remains mostly frozen, except in some local zones (see Fig.~\ref{map} bottom right). We consider a single zone, where each concentric loop with size $l_k \sim k$ can be in two states.  As long as $l_k<l^*$ (this implies that $k=1,\dots,p=L/5$ where $p$ satisfies $l_{p+1}=l^*$) no new loop with $l<l^*$ is generated if a loop flips. There are therefore $2^p$ excited configurations that are specified by a set of $p$ integers $\{ n_k\}$ denoting loops either in their ground state ($n_k=0$) or in an excited state ($n_k=1$).
A single loop excitation with $k>1$ creates two interfaces and costs ``surface'' energy, $E_k \sim k$. Given that the model is short-ranged, there is a short-range interaction between two loops (nearest neighbor), which is an attraction.
 We find no interaction energy beyond two loops, except for the pairwise attraction. The energy of any of these restricted configurations reads
\begin{equation}
E =  \sum_{k=1}^{p} E_k n_k  -  \sum_{k=1}^{p-1} (E_k+E_1) n_k n_{k+1}
\label{model}
\end{equation}
where $E_k=3l_k$ increases with $k$. If we flip all loops ($n_k=1$), we create a new ordered zone (but rotated by 60$^o$) and this costs only ``surface'' energy of order $L$. Since $E_k$ is linear in $k$, the first term has a contribution in $L^2$ (recall that $p \sim L$) which must be canceled by the second term. This explains the general form of the energy, Eq.~(\ref{model}). Note that the highest energy state (with the same diverging $l^*$) consists of flipping every other loop up to $p$. It has a finite excess energy per site of $1/25$.

The problem is the same as that of an Ising chain with $p$ sites and open boundary conditions, with a coupling constant which increases linearly with the bond position. Its partition function can be obtained exactly by transfer matrices (but has not as far as we know) and reads
\begin{equation}
Z = \sum_C e^{-\beta E} = \prod_{k=1}^{p} (1+x^k)=\sum_{k=0}^{p(p+1)/2} p_d(k) x^k 
\end{equation}
where we have expanded the product as a polynomial of $x = \exp(- \beta E_1)$. The coefficients $p_d(k)$ appear to be the number of partitions of the integer $k$ into distinct summands when $p$ is large enough. $p_d(k)$ can be computed exactly and, in the thermodynamic limit ($p \rightarrow \infty$), the asymptotic form reads,\cite{AbramowitzStegun}
\begin{equation}
p_d(k) \sim \frac{1}{k^{3/4}} e^{\pi \sqrt{k/3}}
\end{equation} 
This is the degeneracy of the energy level $E_k=kE_1$, and therefore
its logarithm is 
\begin{equation}
S(E) \equiv \ln p_d(k) \sim \pi \sqrt{\frac{E}{3E_1}}
\end{equation} 
If we assume local thermal equilibrium at temperature $T$, we have 
$\frac{1}{T}=\frac{\partial S}{\partial E}$,
leading to $E(T)=\frac{\pi^2}{4E_1} T^2$ for the three independent zones, hence a linear specific heat
\begin{equation}
C= \frac{\pi^2}{36} T
\label{spe}
\end{equation}
The linear dependence is rather unusual in classical systems, but might apply more generally at grain boundaries where localized excitations are. The important point is a linear increase of both the energy and the interaction for longer excitations that both compensate when a new ordered grain is formed [Eq.~(\ref{model})]. Equation (\ref{spe}) is valid for the metastable ground state at temperatures above the energy gap, $E_1$, when many loops (a few in practice) are excited, and up to infinite temperatures (the ground state is metastable), since it is always possible in the thermodynamic limit to find longer loops. Importantly, note that $C$ is nonextensive and we need a finite density of such zones to obtain an extensive quantity. This is typically obtained in a polycrystalline state (see the density of ``concentric'' structures in Fig.~\ref{map}, bottom left). In this case, the available loop lengths do not extend to infinity, and there is a maximum in the specific heat determined by the longer loop excited, at a given time scale.

\section{Discussion of some analogy with structural glasses}
\label{discussion}

\subsection{General slowing down mechanism} 

We discuss the hypothesis of Adam and Gibbs\cite{AG} in the view
of the present model. We identify the ``cooperatively rearranging
regions'' with the loop degrees of freedom, and the time scale is
activated 
\begin{equation}
\tau_l=\tau_0 \exp(\kappa l/T),
\end{equation} 
for each individual
rearranging region, a point that we discuss below. Each
rearranging region of size $l$ has an entropy 
\begin{equation}
s^*=\ln \Omega
\end{equation}
($\Omega=2$ in the present model). In Adam and Gibbs, the rearranging
regions are \textit{independent} and form $N/l$ domains with $l$ molecules,
so that the total entropy is 
\begin{equation}
S=\frac{N}{l}s^*
\end{equation} 
A direct consequence
of this assumption is that the size $l$ of the smallest
cooperatively rearranging region increases as the entropy decreases
(the domains grow).  

The first essential difference with the present model is that the loop
degrees of freedom are not independent: When a loop flips, it
reorganizes the neighboring loops which may in turn be flippable. A
second difference is that the size of the smallest cooperatively
rearranging regions does not grow here: It is always $l=6$ (down to
the ground state) and these loops do have dynamics. The point is that
they can rearrange some regions locally but may not rearrange the
system globally. Moreover, as the domains of the ordered state grow
(at low energy), the regions insulated from the small-scale loops not
only grow, but also get insulated from larger and larger scales.  This
gives a different interpretation for growing barriers.  Note that the
occurrence of growing domains is not sufficient.  Indeed, in the dimer
model, as we showed, the domains grow but $l^*$ does not, because
these domains are not insulated. The Adam-Gibbs interpretation does
not apply in this case, precisely because there is no way to protect
the ordered domains from the cooperatively rearranging regions (which
are not independent domains).  Adam and Gibbs indeed do not describe
why the dynamics is activated with a relaxation time $\tau_0
\exp(\kappa l/T)$ for an entire domain of $l$ molecules. This is at
odds with the dynamics of the dimer model as explained above, but also
more simply with the Ising model, where rearranging a domain does not
imply an activation energy: The domain walls may move and rearrange
the domain on a fast time scale $l^z$ (where $z$ is a dynamical
exponent). In the context of glasses, a mechanism must then protect
the domain from this fast mechanism.  The present model gives a
solution to this point: No short scales can rearrange the insulated
regions because their borders are protected.  It is therefore not the
number of molecules in the domain that controls the relaxation time,
but the size $l^*$ of the smallest loops that will rearrange the
deepest insulated region.  Moreover, the hierarchical structure of
these regions gives multiple intermediate time scales, a point which 
is also evidenced in structural glasses, for example, from the
stretched exponential behavior of the relaxation functions, and was
phenomenologically described.\cite{Palmer}

\subsection{Microscopic analogy}

The microscopic analogy consists of identifying the chain-like
excitations observed in molecular dynamics simulations of structural
glasses\cite{Kob} with the present loops.  In fact, it was even found
that their averaged size grows slightly on lowering the
temperature,\cite{Kob} just as in the present model. They are
certainly more complicated because of the structurally disordered
environment and their organization in space is not clear.

Pursuing the analogy, we may see (1) the monocrystal as infinite
chains (in two spatial dimensions) that cannot rearrange themselves
(then the crystal is metastable in absence of dislocations); (2) the
supercooled liquid has local collective rearrangements on different
scales because the chains are not structurally ordered, and has
developed hierarchical insulated regions (which are not crystalline
domains); (3) the liquid phase at higher energy has lost the
hierarchical structure and can rearrange itself on the shortest time
scale.  We have distinguished two cases: Either the crystal can be
reached by moving elemental loops of a few molecules (as in the
two-color dimer model) or not (as in the three-color model). This
is reminiscent of the physics of hard discs where the monodisperse
particles can form a solid phase but the polydisperse ones generally
leads to a glassy phase.

The activated behavior of the individual loops $\tau_l=\tau_0
\exp(\kappa l/T)$ is also assumed in the present work, and we have not
specified $\kappa$.  The energy barriers may result from (1) a local
thermodynamic equilibrium: Flipping a loop increases the energy
because of the interactions with the neighboring sites. Alternatively,
(2) it may result from an interaction with the long wavelength
vibrations. If flipping a loop does not cost any energy (for instance,
the local interactions are translationally invariant), such
displacements create nevertheless some diffusing potentials for the
long-wavelength harmonic oscillations, which may result in preferred
positions and barriers between them.\cite{cepascanals} 
An important point is the presence of hard local constraints. If the
constraint is softened (noninfinite nearest-neighbor coupling $J_1$),
it is possible to create a pair of defects that can diffuse fast and
annihilate.\cite{Castelnovo} This is equivalent to flipping the loops
but it occurs typically on time scale $\tau_{d} \sim l^2 \exp
(J_1/T)$. This process is dominant for loops longer than
$l>J_1/\kappa$, giving a cutoff for the diverging barriers. The
activation is therefore relevant only if a strong enough local
constraint prevents the nucleation of such defects. 

We have identified the equilibrium long-range-ordered phase below
$T_m$ with the solid phase since it breaks both the translational and
the orientational symmetries; and the paramagnetic phase with the liquid
which can be supercooled down to $T_{sp}$. The internal energy of the
liquid is well defined down to $T_{sp}$. If we assume that it varies
as $\langle E \rangle(T) \sim E_{\infty} -C/T$, a Vogel-Fulcher law
results from Eq.~(\ref{lstar}), down to $T_{sp}$. In fact, there is only
a discrete increase of the energy barriers ($l^*$ is discrete) and
their enhancement factor, 14/6 at $T_{sp}$, is relatively small in
this model.  Below $T_{sp}$, there is no liquid ``phase'' anymore. The
system is out-of-equilibrium and gets stuck into some basin of energy,
which depends on the cooling rate. We may imagine that if the cooling
is very slow and becomes comparable to $\tau_{d}$, the ground state
may be reached, thanks to the defects. It is an interesting issue and
the equilibration does not always occur either.\cite{Moessner,Levis}

Hard constraints also lead to some features apparently different from
standard liquids.  The paramagnetic phase does not break any symmetry
but has, unlike a true liquid, an infinite correlation length. The
correlations are algebraic and therefore a quasi-Bragg peak is
expected in the structure factor instead of a broad feature.  The
quasi-Bragg peak appears at the $Q$ vector corresponding to the
$\sqrt{3} \times \sqrt{3}$ unit-cell and is at a different position
from that of the solid. It can be seen as a preferred structure in the
liquid phase (minimizing the strong local constraint), and this
structure is not compatible with long-range orientational order, since
it favors short loops.  Again, a softening of the hard constraint will
generate defects and a finite correlation length.

The linear specific heat found in Sec. \ref{excited} cannot be
directly applied to amorphous states at low temperatures, because it
is at temperatures above the energy gap, $T>E_1$, determined by the
smallest loop. Note, however, that the quantum tunneling of the loops
that we have not treated here, or a coupling to the bath of long
wave-length vibrations, may extend its regime of validity to lower
temperatures. Here the model simply assumes some classical interacting
two-level systems, with a physical interpretation. They represent
localized collective excitations that can reconstruct a new ordered
domain locally, and may also be relevant under an external shear.  The slope of the
specific heat is argued to be proportional to the density of
such zones in the limit of small density.
 
\vspace{-0.2cm}
\section{Conclusion}

We have identified a model within which static moments appear when the
scales of the dynamical degrees of freedom are smaller than a length
scale $l^*$ that we have computed.  This length scale grows at low
energy, providing growing energy barriers in the relaxation of the
system. The smaller scales are unable to
reorganize the system globally, leaving some
insulated regions in space. These regions form a hierarchical
structure, and the reorganization of the deeper levels implies multiple time
scales in the dynamics.  Although the system would undergo a
first-order phase transition at thermal equilibrium, we have argued
that it gets trapped in the complex energy landscape that we have
described.

We have discussed a
tentative analogy with structural glasses.  The present
model gives an interpretation of the assumptions made by Adam and
Gibbs to explain the growing barriers.  It suggests in particular that we replace the idea of
independent domains by that of hierarchical insulated regions, the relaxation of
which is controlled by a growing $l^*$ rather than its size.

\vspace{-0.2cm}
\acknowledgements

I would like to thank J.-L. Barrat, B. S. Shastry and T. Ziman for discussions. 
Also, I thank B. S. Shastry for suggesting the term ``geometrically insulated''.


\begin{thebibliography}{99}
\bibitem{effective} This can be seen as an effective
dynamics of the continuous frustrated Heisenberg model,\cite{cepascanals} 
which was conjectured to be glassy.\cite{Chandra}
\end{thebibliography}
\end{document}